\documentclass[twocolumn,prb,showpacs,preprintnumbers,amsmath,amssymb]{revtex4}

\usepackage{graphicx}

\begin{document}


\title{Electronic structure and magnetism of the diluted magnetic semiconductor\\ Fe-doped ZnO nano-particles}

\author{T. Kataoka$^1$, M. Kobayashi$^1$, Y. Sakamoto$^1$, G. S. Song$^1$, 
        A. Fujimori$^{1,2}$, 
        F.-H. Chang$^3$, 
        H.-J. Lin$^3$,\\ 
        D. J. Huang$^3$, 
        C. T. Chen$^3$, 
        T. Ohkochi$^2$, 
        Y. Takeda$^2$, 
        T. Okane$^2$, 
        Y. Saitoh$^2$, 
        H. Yamagami$^{2,4}$,\\ 
        A. Tanaka$^5$, 
        S. K. Mandal$^6$, 
        T. K. Nath$^6$, 
        D. Karmakar$^{7,8}$, 
       and I. Dasgupta$^9$}
\affiliation{$^1$Department of Physics and Department of Complexity Science and Engineering, 
University of Tokyo, Bunkyo-ku, Tokyo 113-0033, Japan\\
$^2$Synchrotron Radiation Research Unit, Japan Atomic Energy Agency,
Sayo-gun, Hyogo 679-5148, Japan\\
$^3$National Synchrotron Radiation Research Center, Hsinchu 30076, Taiwan\\
$^4$Department of Physics, Faculty of Science, Kyoto Sangyo University, Kyoto 603-8555, Japan\\
$^5$Department of Quantum Matter, ADSM, Hiroshima University, Higashi-Hiroshima 739-8530, Japan\\
$^6$Department of Physics \& Meteorology, Indian Institute of Technology, Kharagpur 721302, India\\
$^7$Technical Physics \& Prototype Engineering Division, Bhabha Atomic Research Center, Mumbai 400085, India\\
$^8$Present address: Max Planck Institut Fur Festkorperforschung, Heisenbergstrasse 1, Stuttgart D-70569, Germany\\
$^9$Department of Solid State Physics and Center for Advanced Materials, 
Indian Association for the Cultivation of Science, Jadavpur Kolkata 700032, India}

\date{\today}

\begin{abstract}
We have studied the electronic structure of Zn$_{0.9}$Fe$_{0.1}$O nano-particles, 
which have been reported to show ferromagnetism at room temperature, 
by x-ray photoemission spectroscopy (XPS), resonant photoemission spectroscopy (RPES), 
x-ray absorption spectroscopy (XAS) and x-ray magnetic circular dichroism (XMCD). 
From the experimental and cluster-model calculation results, 
we find that Fe atoms are predominantly in the Fe$^{3+}$ ionic state with mixture of a small amount of Fe$^{2+}$ and 
that Fe$^{3+}$ ions are dominant in the surface region of the nano-particles. 
It is shown that the room temperature ferromagnetism in the Zn$_{0.9}$Fe$_{0.1}$O nano-particles 
is primarily originated from the antiferromagnetic coupling 
between unequal amounts of Fe$^{3+}$ ions occupying two sets of nonequivalent 
positions in the region of the XMCD probing depth of $\sim$ 2-3 nm.

\end{abstract}


\maketitle
\section{Introduction}
There is growing interest in diluted magnetic semiconductors (DMSs), 
where magnetic ions are doped into the semiconductor hosts, 
due to the possibility of utilizing both charge and spin degrees of 
freedom in the same materials, allowing us to design a new generation 
spin electronic devices with enhanced functionalities \cite{Furdyna, Ohno}. 
Theoretical studies on the basis of Zener's $p$-$d$ exchange model have shown 
that wide-gap semiconductors such as ZnO doped with transition metal are 
promising candidates for room temperature ferromagnetic DMSs \cite{Dietl}. 
First-principle calculations by Sato and Katayama-Yoshida \cite{Katayama} 
have also predicted that ZnO-based DMSs exhibit ferromagnetism using LSDA calculation. 
Subsequently a number of experiments on ZnO-based DMSs in bulk, 
thin film and nano-particle forms revealed ferromagnetic properties 
\cite{Tuan, Ueda, KVRao, Schwartz, Radovanovic, Melon}, 
and among them ZnO-based DMSs nano-particles have attracted much attention \cite{Mandal1, Mandal2}. 
Current interest in such magnetic nano-particle systems is motivated by 
unique phenomena such as superparamagnetism \cite{Kittle}, quantum tunneling of magnetization \cite{Gunther} 
and particularly magnetism induced by surface effects \cite{Garcia}. 
In the nano-particle form, the structural and electronic properties are modified 
at the surface as a result of the broken translational symmetry of the lattice or 
dangling bond formation, giving rise to weakened exchange coupling, 
site-specific surface anisotropy, and surface spin disorder \cite{Sun, Kodama}. 
That is, the modification of the electronic structure at the surface of the nano-particles 
plays a crucial role in the magnetism of this system. 

Recently, Karmakar ${et}$ ${al}$. \cite{Karmakar} have reported room temperature ferromagnetism 
in Fe-doped ZnO nano-particles in the proposed core/shell structure, 
where Fe$^{2+}$ ions are situated mostly in the core and Fe$^{3+}$ ions in the surface region. 
However, LSDA+U calculation \cite{Gopal} has indicated the insulating antiferromagnetic 
state to be more stable than the ferromagnetic state for Fe-doped ZnO system. 
In view of the presence of Fe$^{3+}$ ions as indicated by local magnetic probes such as 
electron paramagnetic resonance (EPR) and M\"ossbauer measurements, 
Karmakar ${et}$ ${al}$. \cite{Karmakar} have proposed that the presence of surface Zn vacancies that 
dope hole into the system will be more effective to stabilize the ferromagnetism in this system. 
However, the correlation between magnetic properties and electronic structure of the 
Fe-doped ZnO nano-particle semiconductors has not been clarified yet. 
Thus, investigation of the electronic structure of the Fe-doped ZnO nano-particles 
is critical to achieve better understanding of this type of nano-materials and 
to perform new material design. 
In this paper, we have investigated the electronic structure of 
Zn$_{0.9}$Fe$_{0.1}$O nano-particles 
using x-ray photoemission spectroscopy (XPS), 
vacuum ultraviolet and soft-x-ray resonant photoemission spectroscopy (RPES), 
x-ray absorption spectroscopy (XAS) and x-ray magnetic circular dichroism (XMCD). 
RPES is a convenient tool to obtain the Fe 3$d$ partial density of states (PDOS) in the valence-band spectra \cite{Davis}. 
By performing RPES in the Fe 3$p$-3$d$ core-excitation region, 
we have studied the electronic states 
in the surface region with a probing depth of $\sim$ 0.5 nm \cite{Lindau} 
of the nano-particles utilizing the surface sensitivity of the technique. 
On the other hand, RPES in the Fe 2$p$-3$d$ core-excitation region 
is more bulk sensitive with a probing depth of $\sim$ 1.5-2.0 nm \cite{Lindau} 
and enables us to study the electronic structure in both 
the core and surface regions of the nano-particles. 
XAS and XMCD, whose probing depth are $\sim$ 2-3 nm, 
enable us to study the element specific electronic structure of the Zn$_{0.9}$Fe$_{0.1}$O nano-particles. 
In particular, XMCD is a powerful tool to study element-specific local magnetic states. 
Based on our experimental results, we shall discuss the origins of the ferromagnetic properties and 
magnetic interactions in the Zn$_{0.9}$Fe$_{0.1}$O nano-particles.

\section{Experiment}

Zn$_{0.9}$Fe$_{0.1}$O nano-particles were prepared by the chemical pyrophoric reaction technique. 
Structural characterization was carried out using x-ray diffraction and transmission electron microscopy (TEM), 
demonstrating a clear nano-crystal phase. As observed by TEM, the average particle size was around 7 nm with 
the particle size distribution of 3.0-30.0 nm. 
Magnetization measurements on the same samples revealed a ferromagnetic-to-paramagnetic 
transition temperature $>$ 450 K. Details of the sample preparation were described in Ref. [18]. 
We measured a pressed pellet sample, which after grinding had been calcined at 350Ž. 
The ferromagnetic moment per Fe as deduced from 
the SQUID magnetization data was $\sim$ 0.05 $\mu$$_B$ at room temperature \cite{Karmakar}. 
XAS and XMCD measurements were performed at the Dragon Beamline BL11A of 
National Synchrotron Radiation Research Center (NSRRC) in the total-electron-yield (TEY) mode. 
The monochromator resolution was $E$$/$$\Delta$$E$ $>$ 10000 and the circular polarization of x-rays was $\sim$ 55 \%. 
XPS measurements using the photon energy of $h$$\nu$ = 1253.6 eV were performed at BL23-SU of SPring-8. 
RPES measurements in the Fe 2$p$-3$d$ and 3$p$-3$d$ core-excitation regions were performed at BL23-SU of SPring-8 
and at BL-18A of Photon Factory (PF), respectively. 
For the photoemission measurements, 
all binding energies ($E$$_B$) were referenced to the Fermi level ($E$$_F$) of the 
sample holder which was in electrical contact with the sample. 
The total energy resolutions of the XPS and RPES measurements 
were $\sim$ 400 meV and $\sim$ 170 meV, respectively. 
All the experiments were performed at room temperature.

\section{Results and discussion}
\begin{figure}
\includegraphics[width=4.5cm]{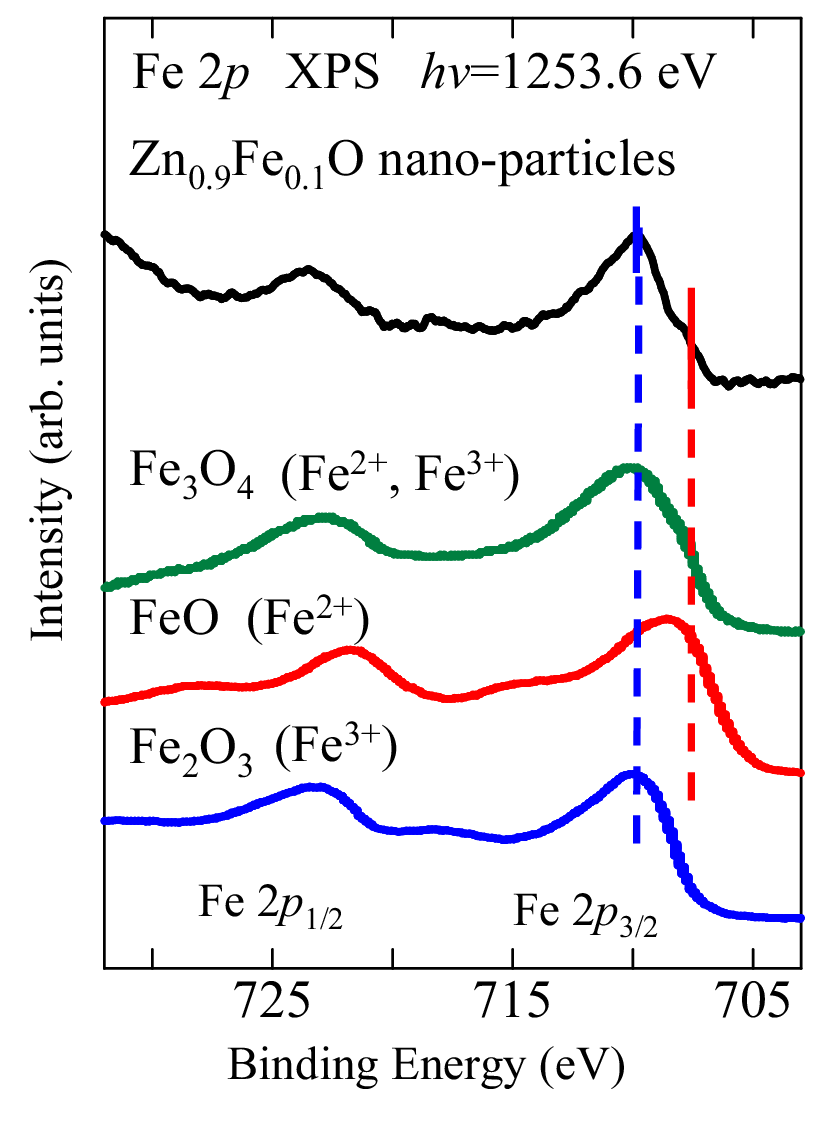}
\caption{\label{XPS}(Color online) 
Fe 2$p$ core-level XPS spectrum of the Zn$_{0.9}$Fe$_{0.1}$O nano-particles compared with 
the XPS spectra of $\alpha$-Fe$_2$O$_3$ \cite{Graat}, FeO \cite{Graat} and Fe$_3$O$_4$ \cite{Fujii}.} \end{figure}

Figure \ref{XPS} shows the Fe 2$p$ core-level XPS spectrum of the Zn$_{0.9}$Fe$_{0.1}$O 
nano-particles in comparison with those of $\alpha$-Fe$_2$O$_3$ (Fe$^{3+}$) \cite{Graat}, 
FeO (Fe$^{2+}$) \cite{Graat} and Fe$_3$O$_4$ (Fe$^{3+}$-Fe$^{2+}$ mixed-valence) \cite{Fujii}. 
The Fe 2$p$$_{3/2}$ peak of the Zn$_{0.9}$Fe$_{0.1}$O nano-particles is split into two peaks at $E$$_B$ $\sim$ 710 eV and $\sim$ 708 eV, 
corresponding to the energy positions of $\alpha$-Fe$_2$O$_3$ and FeO. 
The XPS spectrum of the Zn$_{0.9}$Fe$_{0.1}$O nano-particles therefore reflects 
an Fe$^{3+}$-Fe$^{2+}$ mixed-valent state of the Fe ions in agreement with the 
previous M\"ossbauer report \cite{Karmakar}. 
\begin{figure}
\includegraphics[width=7.0cm]{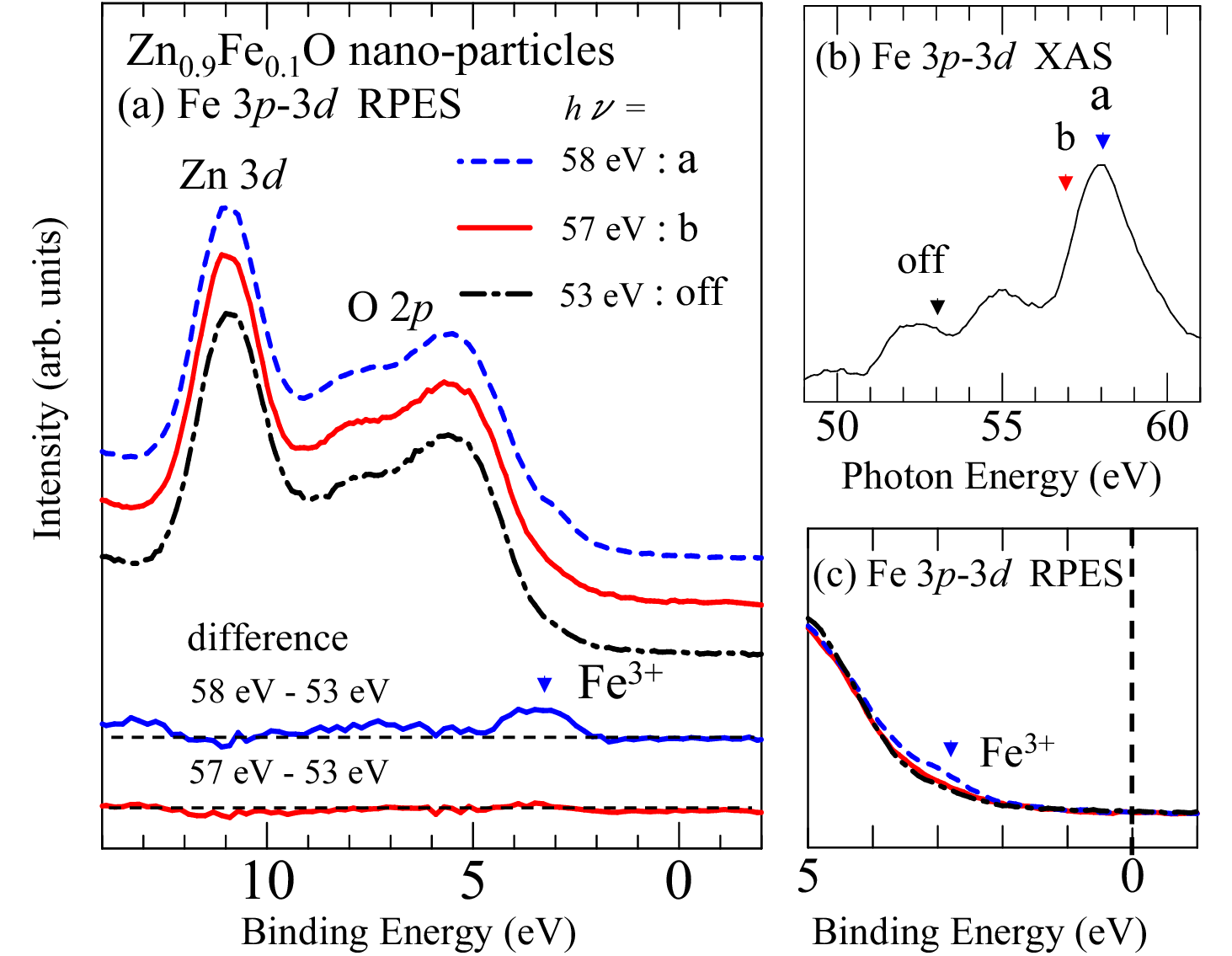}
\caption{\label{3p-3dRPES}(Color online) Valence-band photoemission spectra of the Zn$_{0.9}$Fe$_{0.1}$O nano-particles. 
(a) A series of photoemission spectra taken with photon energies in the Fe 3$p$-3$d$ core-excitation region. 
Difference curves at the bottom represent the Fe 3$d$ PDOS. 
(b) Fe 3$p$-3$d$ XAS spectrum. 
(c) Magnified view near the valence-band maximum.} \end{figure}
In a Fe-doped ZnO system, the valence state of Fe is expected to be +2 if Fe simply substitutes for Zn. 
The presence of Fe$^{3+}$ ions in this sample has been suggested due to surface Zn vacancies \cite{Karmakar} or 
excess oxygens of the nano-particles.

\begin{figure}
\includegraphics[width=8.0cm]{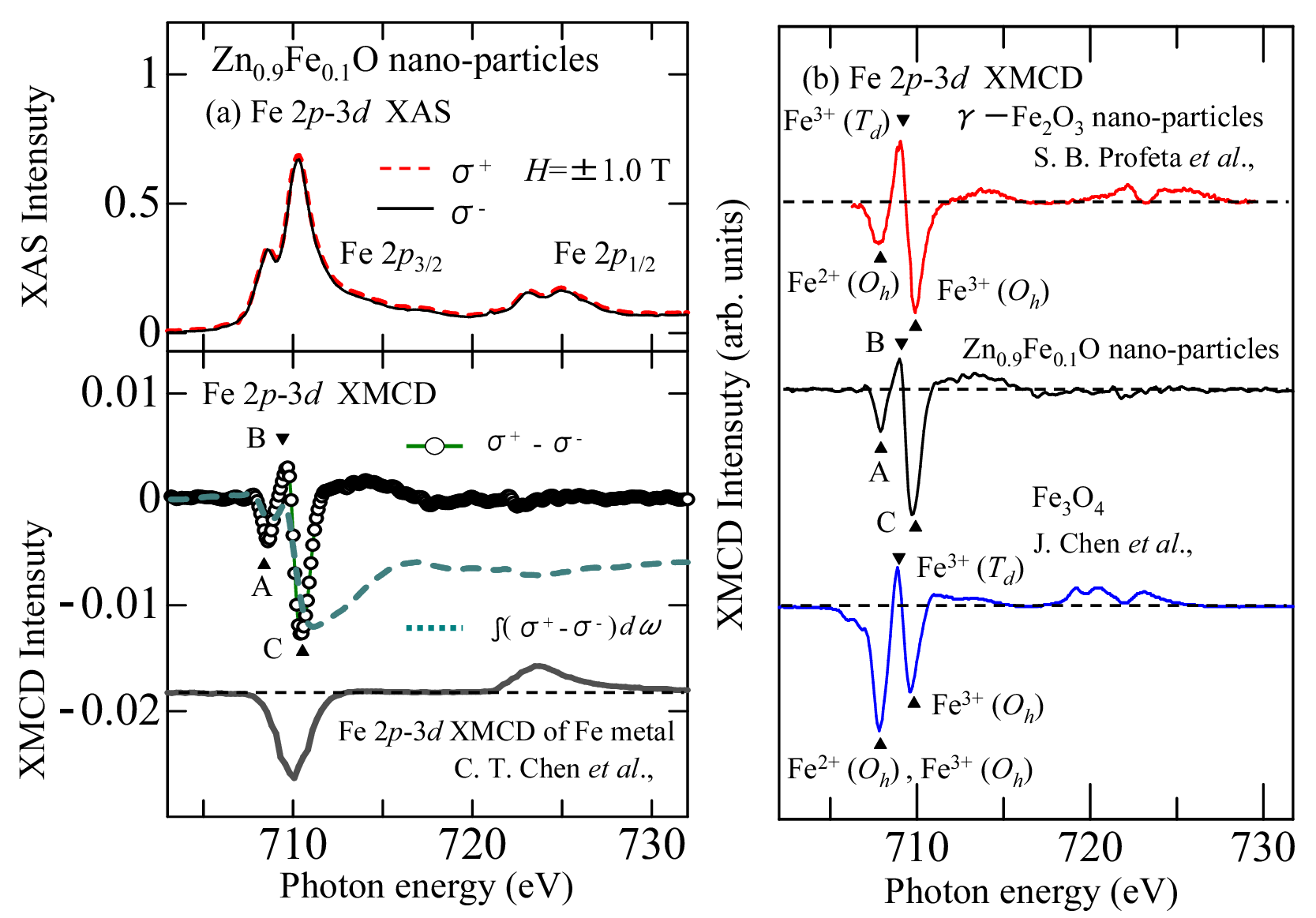}
\caption{\label{XMCD}(Color online) Fe 2$p$-3$d$ XAS and XMCD spectra of the Zn$_{0.9}$Fe$_{0.1}$O nano-particles 
compared with those of other Fe oxides. 
(a) Fe 2$p$-3$d$ XAS spectra in magnetic fields of $\pm$1 T. 
XAS ($\sigma$$^+$ and $\sigma$$^-$), XMCD and its integral. 
The XMCD spectrum of Fe metal is shown for comparison \cite{Chen}. 
(b) XMCD spectrum of the Zn$_{0.9}$Fe$_{0.1}$O nanoparticles compared with the XMCD spectra of 
$\gamma$-Fe$_2$O$_3$ \cite{Profeta} and Fe$_3$O$_4$ \cite{JChen}.} \end{figure}
In order to study the electronic states in the surface region of the nano-particles, 
we performed RPES measurements in the Fe 3$p$-3$d$ core-excitation region. 
RPES in 3$d$ transition metals and their compounds are caused by interference between 
direct photoemission from the 3$d$ level and Auger-electron emission following the 3$p$ (2$p$)-3$d$ core-excitation \cite{Davis}. 
Therefore, the difference between valence-band spectra measured on- and off-resonance 
is used to extract the resonantly enhanced Fe 3$d$ contributions to the valence-band region. 
Figure \ref{3p-3dRPES}(a) shows the valence-band photoemission spectra of the Zn$_{0.9}$Fe$_{0.1}$O nano-particles 
taken with various photon energies in the Fe 3$p$-3$d$ core-excitation region 
marked on the Fe 3$p$-3$d$ XAS spectrum [Fig. \ref{3p-3dRPES}(b)]. 
Figure \ref{3p-3dRPES}(c) shows a magnified view near the valence-band maximum. 
In Fig. \ref{3p-3dRPES}(b), one can see that a peak appears at 58 eV, representing the Fe 3$p$-3$d$ absorption. 
The same peak is found at 58 eV for $\alpha$-Fe$_2$O$_3$ (Fe$^{3+}$) \cite{Lad}. 
For FeO (Fe$^{2+}$), on- and off-resonance energies are reported to be 57 and 53 eV, respectively \cite{Fujimori}. 
From this comparison, we conclude that 3$p$-3$d$ absorption is mainly due to Fe$^{3+}$ ions. 
In Fig. \ref{3p-3dRPES}(c), one can see that in going from the off-resonance spectrum ($h$$\nu$ = 53 eV) to 58 eV, 
the tale at $E$$_B$ $\sim$ 3-4 eV grows in intensity. 
By subtracting the off-resonance spectrum from the on-resonant ones of 
Fe$^{3+}$ ($h$$\nu$ = 58 eV) and Fe$^{2+}$ ($h$$\nu$ = 57 eV), respectively, 
we have extracted the Fe 3$d$ partial density of states (PDOS) of Fe$^{3+}$ and Fe$^{2+}$ 
as shown in the bottom panel of Fig. \ref{3p-3dRPES}(a). 
The Fe$^{3+}$ 3$d$ PDOS reveals a feature at $E$$_B$ $\sim$ 3-4 eV. 
On the other hand, the Fe$^{2+}$ 3$d$ PDOS reveals no clear feature. 
We therefore conclude that the Fe$^{3+}$ ions are dominant in the surface region of 
the Zn$_{0.9}$Fe$_{0.1}$O nano-particles probed by Fe 3$p$-3$d$ RPES. 
This, together with the bulk-sensitive Fe 2$p$-3$d$ RPES result described below, 
may support the core/shell model of the Zn$_{0.9}$Fe$_{0.1}$O nano-particles proposed in the previous report \cite{Karmakar}.

\begin{figure}
\includegraphics[width=6.5cm]{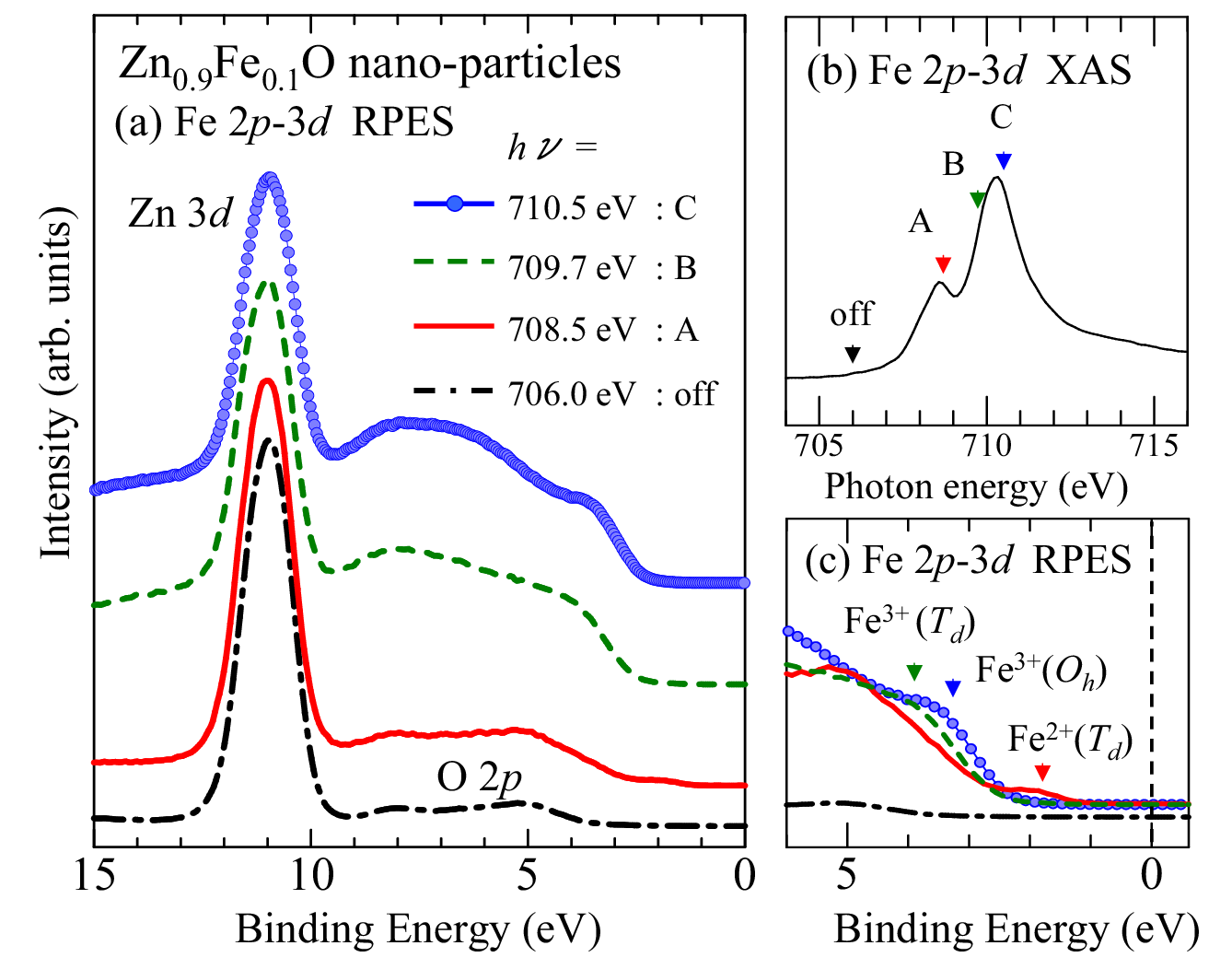}
\caption{\label{2p-3dRPES}(Color online) Valence-band photoemission spectra of the Zn$_{0.9}$Fe$_{0.1}$O nano-particles. 
(a) A series of photoemission spectra for photon energies in the Fe 2$p$-3$d$ core-excitation region. 
(b) Fe 2$p$-3$d$ XAS spectrum. 
(c) Magnified view near the valence-band maximum.} \end{figure}

Figure \ref{XMCD}(a) shows the Fe 2$p$-3$d$ XAS and XMCD spectra of the Zn$_{0.9}$Fe$_{0.1}$O  
nano-particles for opposite magnetization directions recorded using circular polarized x-rays, 
their difference spectrum, i.e., XMCD spectrum, and its integration. 
Here, the XAS spectra obtained in the magnetic field of +1.0 T and -1.0 T are denoted by $\sigma$$^+$ and $\sigma$$^-$, respectively. 
The bottom panel shows the XMCD spectrum of Fe metal \cite{Chen}. 
In the XMCD spectrum of the nano-particles, 
three sharp peaks around $h\nu$ = 708.5, 709.7 and 710.5 eV, denoted by A, B and C, respectively, are observed. 
The XMCD spectral line shape of the Zn$_{0.9}$Fe$_{0.1}$O  nano-particles is different from that of Fe metal, 
indicating that the magnetism in this sample is not due to segregation of metallic Fe clusters but due to 
the ionic Fe atoms with localized 3$d$ electrons.

Figure \ref{XMCD}(b) shows the Fe 2$p$-3$d$ XMCD spectrum of the Zn$_{0.9}$Fe$_{0.1}$O nano-particles 
in comparison with those of $\gamma$-Fe$_2$O$_3$ nano-particles, where Fe$^{3+}$ ions are both at the 
tetrahedral ($T$$_d$) and octahedral ($O$$_h$) sites \cite{Profeta}, 
and Fe$_3$O$_4$, where Fe$^{3+}$ ions at the $T$$_d$ and $O$$_h$ sites and Fe$^{2+}$ ions at the $O$$_h$ sites coexist \cite{JChen}. 
The XMCD spectrum of the Fe$_3$O$_4$, which displays the overlapping contributions from the Fe$^{3+}$ and Fe$^{2+}$ ions, 
is different from that of the Zn$_{0.9}$Fe$_{0.1}$O nano-particles. 
On the other hand, the spectral line shape of the $\gamma$-Fe$_2$O$_3$ nano-particles, 
where XMCD signals are due to Fe$^{3+}$, is similar to that of the Zn$_{0.9}$Fe$_{0.1}$O nano-particles. 
This indicates that the magnetism of the Zn$_{0.9}$Fe$_{0.1}$O nano-particles is originated mainly 
from Fe$^{3+}$ ions and contribution from Fe$^{2+}$ ions appears to be small. 
By comparison with the Fe 2$p$-3$d$ XMCD spectral shape of the $\gamma$-Fe$_2$O$_3$ nano-particles, 
peaks B and C for the Zn$_{0.9}$Fe$_{0.1}$O nano-particles 
are assigned to Fe$^{3+}$ ions at the $T$$_d$ and $O$$_h$ sites, respectively. 
Although it is likely that peak A arises mainly from Fe$^{3+}$ ($O$$_h$) ions, 
the present Fe 2$p$-3$d$ RPES result, which is described below, suggests that peak A may be attributed not only to 
Fe$^{3+}$ ($O$$_h$) but also to a small amount of Fe$^{2+}$ ($T$$_d$) ions.

Figure \ref{2p-3dRPES}(a) shows the valence-band photoemission spectra of the Zn$_{0.9}$Fe$_{0.1}$O nano-particles 
taken with various photon energies in the Fe 2$p$-3$d$ core-excitation region. 
The Fe 2$p$-3$d$ XAS spectrum in the same energy region is shown in Fig. \ref{2p-3dRPES}(b). 
The photoemission spectra were taken using photon energies denoted by A, B and C in Fig. \ref{XMCD} and Fig. \ref{2p-3dRPES}(b) 
and at off-resonance ($h$$\nu$ = 706.0 eV). 
For all the spectra, no photoemission intensity was observed at $E_F$, indicating the localized nature of the carriers. 
The off-resonant spectrum of the Zn$_{0.9}$Fe$_{0.1}$O nano-particles is similar to that of 
ZnO showing a sharp peak at about $E$$_B$ $\sim$ 11.0 eV due to the Zn 3$d$ states as well as a broad feature 
at $E$$_B$ $\sim$ 4.0-9.0 eV due to the O 2$p$ band \cite{Gabas}. 
\begin{figure}
\includegraphics[width=4.5cm]{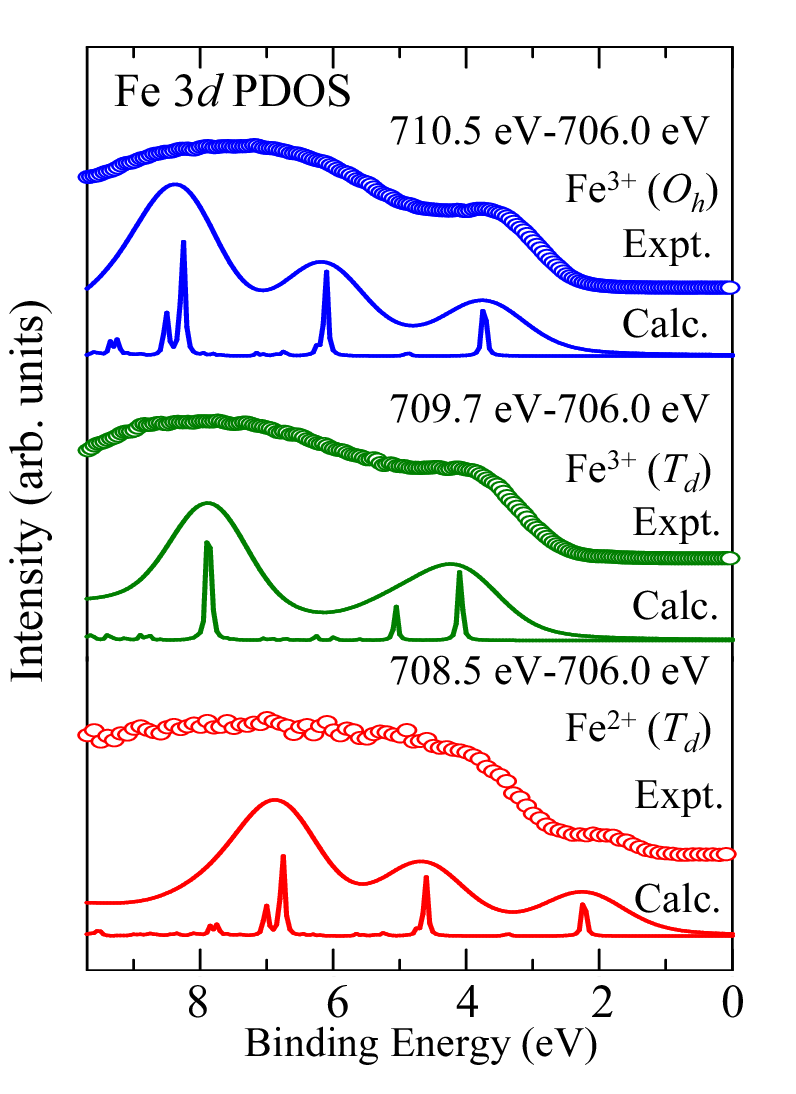}
\caption{\label{PDOS}(Color online) Differences between the on-resonance spectra 
taken with photon energies marked by A, B and C in Fig. \ref{2p-3dRPES}(b) and 
off-resonance one corresponding to the partial density of states of 
the Fe ion in each valence state and crystallographic sites. 
Open circles represent experimental data and thin solid curves indicate calculated spectra.}
\end{figure}
If the photon energy is tuned to peak A ($h$$\nu$ = 708.5 eV), 
one can see a feature at $E$$_B$ $\sim$ 2.0 eV in the on-resonant spectrum [Fig. \ref{2p-3dRPES}(c)], indicating that the intensity of 
photoelectrons arising from Fe$^{2+}$ ions (the minority-spin state of Fe$^{2+}$ ions \cite{Kobayashi}) is enhanced. 
For the spectra excited by photons with energies corresponding to peaks B ($h$$\nu$ = 709.7 eV) and C ($h$$\nu$ = 710.5 eV), 
one can see a broad structure around $E$$_B$ $\sim$ 3.0-9.0 eV [Fig. \ref{2p-3dRPES}(a) and (c)]. 
The spectral line shapes excited by photons corresponding to B and C are similar to each other, 
indicating that on-resonant spectral line shape strongly depends on the valency of Fe ions rather than coordination as anticipated. 
To clarify the electronic structure associated with the Fe 3$d$ ion in each valence state and 
crystallographic site in the Zn$_{0.9}$Fe$_{0.1}$O nano-particles, 
we have performed configuration-interaction cluster-model calculations to deduce the Fe 3$d$ PDOS of each component \cite{Tanaka1, Tanaka2}.

\begin{table}
\caption{\label{parameter} Electronic structure parameters for the Zn$_{0.9}$Fe$_{0.1}$O nano-particles 
used in the cluster-model calculations in units of eV. 
The charge-transfer energy $\Delta$, the on-site 3$d$-3$d$ Coulomb energy $U$$_{dd}$, 
and the 3$d$-2$p$ Coulomb energy $U$$_{dc}$ on the Fe ion, 
the hybridization strength between Fe 3$d$ and O 2$p$ $p$$d$$\sigma$, and the crystal field 10$D$$q$.}
\includegraphics[width=6.0cm]{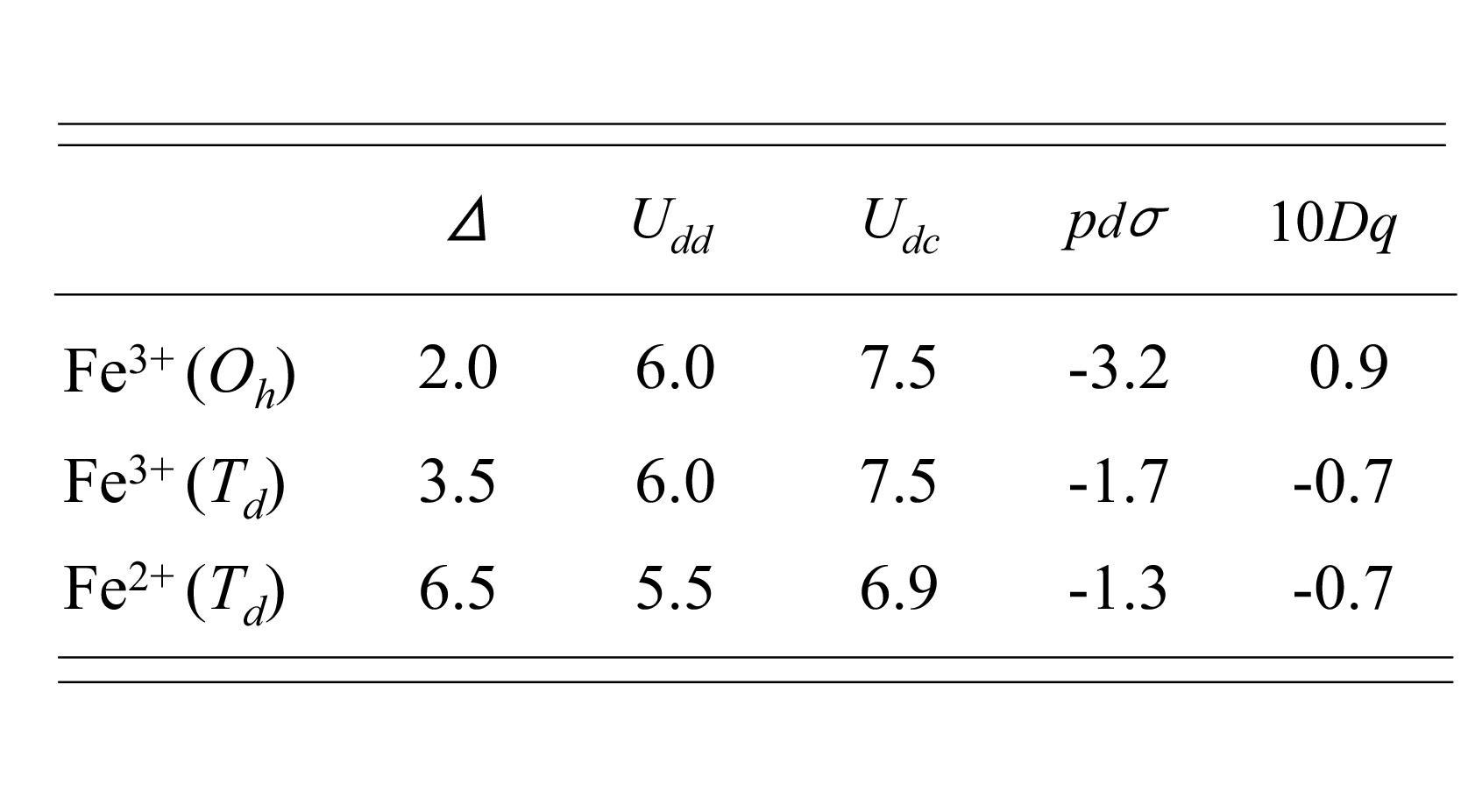}
\end{table}

Figure \ref{PDOS} shows the Fe 3$d$ PDOS (open circles) of the Zn$_{0.9}$Fe$_{0.1}$O nano-particles, 
which has been obtained by subtracting the off-resonance spectrum from the on-resonance 
ones of Fe$^{3+}$ ($O$$_h$), Fe$^{3+}$ ($T$$_d$) and Fe$^{2+}$ ($T$$_d$), respectively. 
Calculated spectra (solid curves) are also shown in the same figure. 
Electronic structure parameters used in the calculations are listed in Table \ref{parameter}. 
For the Fe$^{3+}$ ($T$$_d$ and $O$$_h$ sites) ions in the Fe-doped ZnO nano-particles, 
the values of the on-site 3$d$-3$d$ Coulomb energy $U$$_{dd}$ and the 3$d$-2$p$ Coulomb energy $U$$_{dc}$ 
on the Fe ion have been taken from the literature on the photoemission study of 
Fe$_3$O$_4$ \cite{JChen}, where Fe$^{3+}$ ions at the $T$$_d$ and $O$$_h$ sites and Fe$^{2+}$ ions at the $O$$_h$ sites coexist. 
In addition to this, based on the RPES results, we have chosen reasonable values for 
the charge-transfer energy $\Delta$ of the Fe$^{3+}$ ($T$$_d$ and $O$$_h$ sites) ions. 
The electronic structure parameters ($U$$_{dd}$, $U$$_{dc}$ and $\Delta$) of the Fe$^{2+}$ ($T$$_d$) 
ions were appropriately chosen to reproduce the RPES result since there is no information 
in the literature about the electronic structure parameters of the Fe$^{2+}$ ($T$$_d$) ions. 
The $\Delta$ value of the Fe$^{2+}$ ($T$$_d$) ions thus employed is large compared to those of the Fe$^{3+}$ ($T$$_d$) ions, 
consistent with the systematic decrease in the $\Delta$ value as the ionic charge increases \cite{Bocquet}. 
The calculated spectra have been broadened with a Gaussian having a full width at half maximum (FWHM) of 0.6 eV and 
with a Lorentzian having a FWHM of 0.2 eV. 
The spectral line shapes of the calculated results agree with those of experimental results, 
confirming the presence of the Fe$^{3+}$ ($O$$_h$), Fe$^{3+}$ ($T$$_d$) and Fe$^{2+}$ ($T$$_d$) ions.

\begin{figure}
\includegraphics[width=7.4cm]{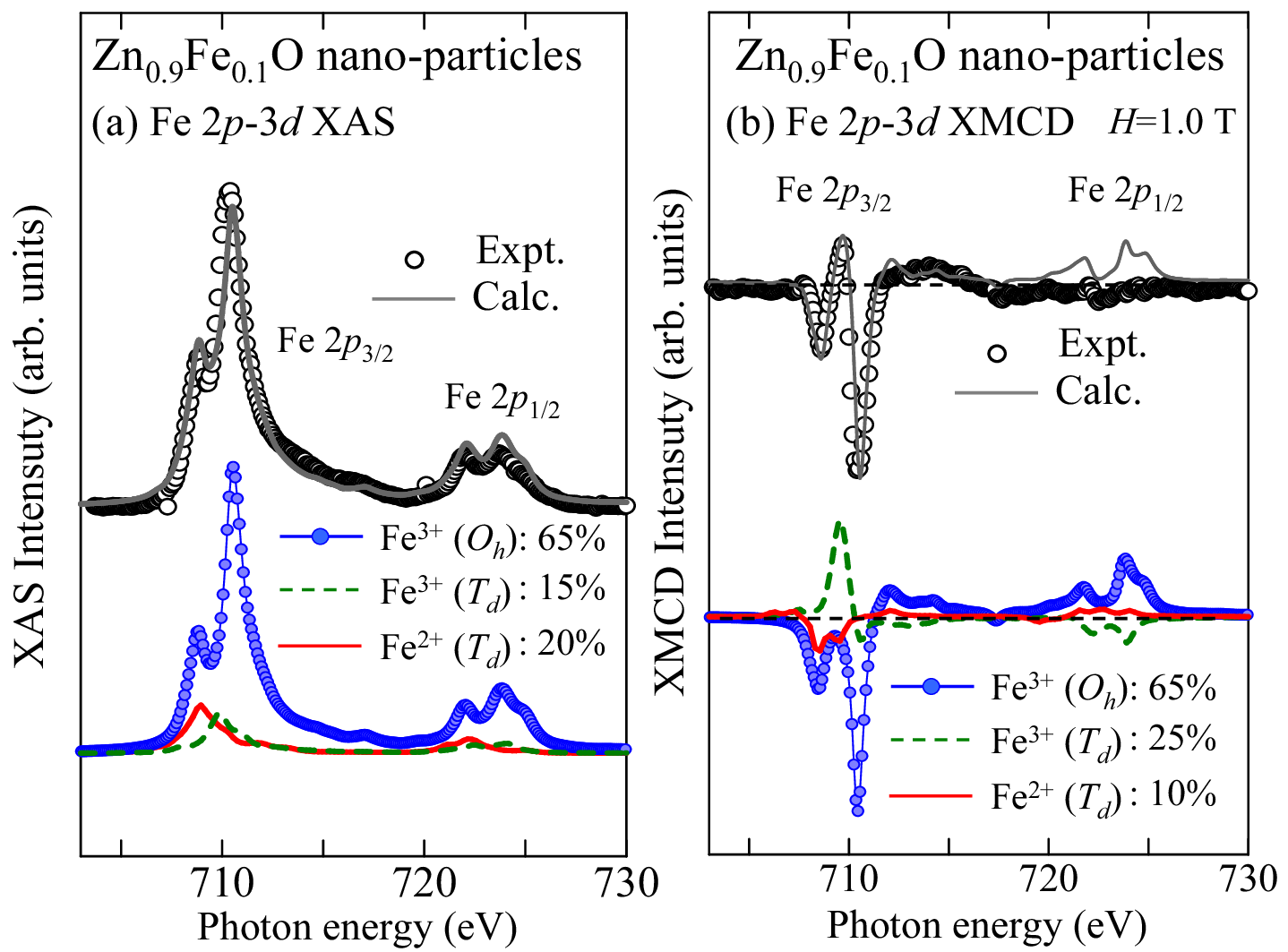}
\caption{\label{Calculation}(Color online) Fe 2$p$-3$d$ XAS (a) and XMCD (b) spectra of 
the Zn$_{0.9}$Fe$_{0.1}$O nano-particles compared with the calculated one using the cluster-model. 
Calculated spectra of the Fe$^{3+}$ ($O$$_h$), Fe$^{3+}$ ($T$$_d$) and Fe$^{2+}$ ($T$$_d$) 
ions at the bottom of each panel have been added to be compared with experiment. 
Parameters used in the calculations (Table \ref{parameter}) were obtained from the analysis of the Fe 2$p$-3$d$ RPES spectra.}
\end{figure}

Figure \ref{Calculation}(a) and (b) shows the Fe 2$p$-3$d$ XAS and XMCD spectra of the Zn$_{0.9}$Fe$_{0.1}$O nano-particles 
compared with the calculated spectra of the Fe$^{3+}$ ($O$$_h$), Fe$^{3+}$ ($T$$_d$) and Fe$^{2+}$ ($T$$_d$) ions. 
The calculations have been made using parameters listed in Table \ref{parameter}. 
One observes shifts of the peaks in the XAS and XMCD spectra for the three kinds of the Fe ions, 
Fe$^{3+}$ ($T$$_d$), Fe$^{3+}$ ($O$$_h$) and Fe$^{2+}$ ($T$$_d$) ions. 
The center of gravity of each spectrum is affected by Madelung energy, $U$$_{dd}$ and $U$$_{dc}$ at the Fe site, 
whereas the peak position may be shifted by crystal-field splitting \cite{JChen}. 
Therefore, both the coordination and the valence state of the Fe ion affect the XAS and XMCD peak positions. 
Thus one can clearly distinguish between the valence and crystal-field of Fe ion [Fe$^{3+}$ ($T$$_d$), 
Fe$^{3+}$ ($O$$_h$) and Fe$^{2+}$ ($T$$_d$)]. 
In Fig. \ref{Calculation}(a), 
the weighted sum of the calculated Fe$^{3+}$ ($O$$_h$: $\sim$ 65 \%), Fe$^{3+}$ ($T$$_d$: $\sim$ 15 \%) and Fe$^{2+}$ ($T$$_d$: $\sim$ 20 \%) 
XAS spectra shown at the bottom of panel (a) approximately reproduces the measured XAS spectrum. 
These ratios indicate that Fe ions are predominantly in the Fe$^{3+}$ state with mixture of a small amount of Fe$^{2+}$. 
In Fig. \ref{Calculation}(b), the weighted sum of the calculated 
Fe$^{3+}$ ($O$$_h$: $\sim$ 65 \%), Fe$^{3+}$ ($T$$_d$: $\sim$ 25 \%) and Fe$^{2+}$ ($T$$_d$: $\sim$ 10 \%) XMCD 
spectra shown at the bottom of panel (b) approximately reproduces the measured XMCD spectrum.

\begin{figure}
\includegraphics[width=8.8cm]{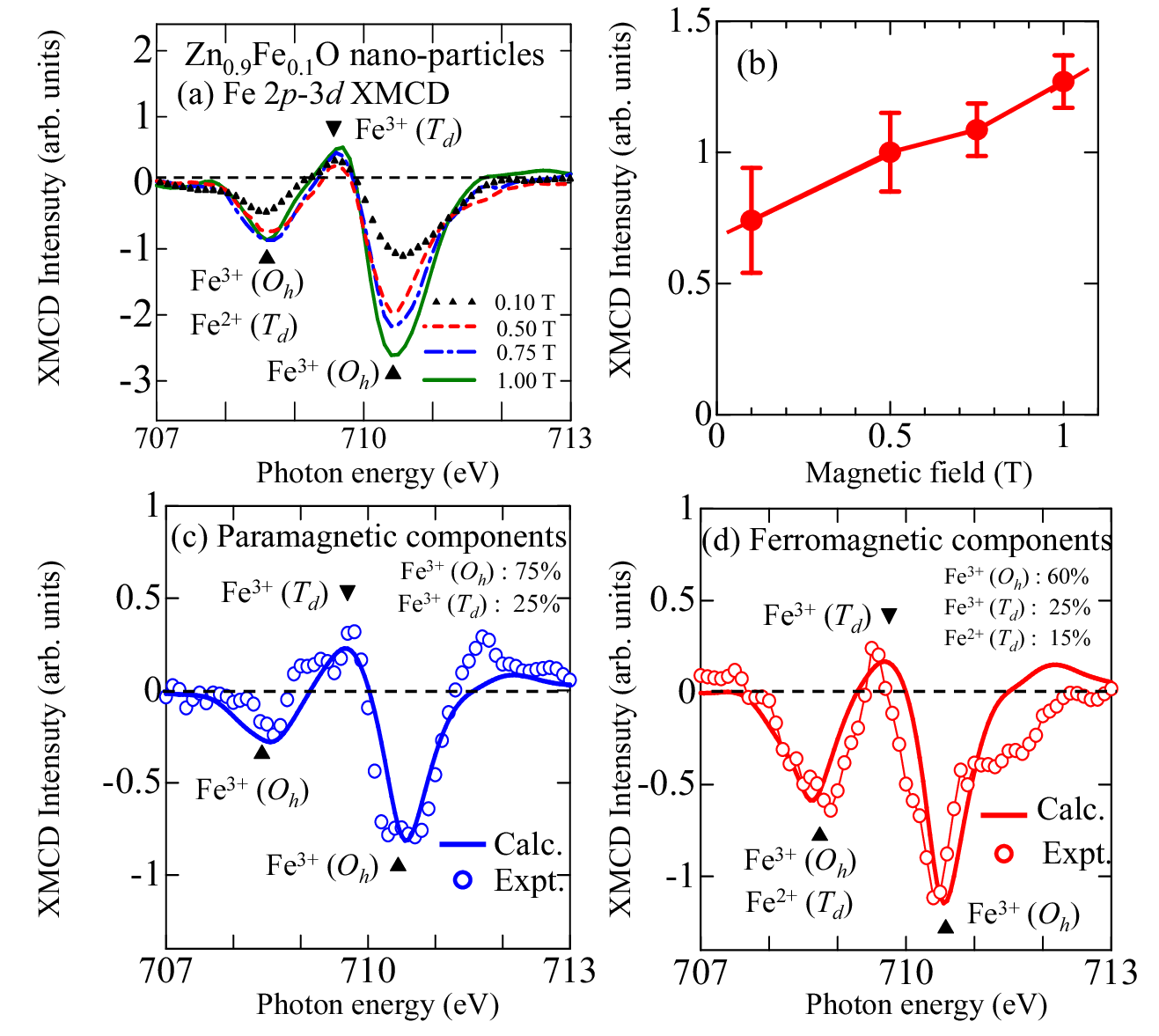}
\caption{\label{MDXMCD}(Color online) Fe 2$p$-3$d$ XMCD spectra of the Zn$_{0.9}$Fe$_{0.1}$O nano-particles. 
(a) Fe 2$p$-3$d$ XMCD spectra of the Zn$_{0.9}$Fe$_{0.1}$O nano-particles measured at various magnetic fields. 
(b) Intensity of Fe 2$p$-3$d$ XMCD (at $h\nu$ = 710.5 eV) as a function of magnetic field. 
Paramagnetic (c) and ferromagnetic (d) components obtaind from the XMCD spectra at 1.0 and 0.5 T. 
Open circles represent experimental data and thin solid curves indicate calculated spectra.
}
\end{figure}

Figure \ref{MDXMCD}(a) shows the Fe 2$p$-3$d$ XMCD spectra of the Zn$_{0.9}$Fe$_{0.1}$O nano-particles measured at various magnetic fields. 
One can observe the XMCD intensity down to $H$ $\sim$ 0.1 T as shown in Fig. \ref{MDXMCD}(a) and (b), 
indicating that the ferromagnetism in this sample is originated from the ionic Fe atoms. 
The difference between the XMCD spectra at $H$ = 1.0 and 0.5 T reflects the paramagnetic components as shown in Fig. \ref{MDXMCD}(c). 
From the line shape of the paramagnetic components analysed with the cluster-model calculation, 
we conclude that the paramagnetism in the Zn$_{0.9}$Fe$_{0.1}$O nano-particles is originated from 
the Fe$^{3+}$ ions ($O$$_h$: $\sim$ 75 \% and $T$$_d$: $\sim$ 25 \%) and contributions from the Fe$^{2+}$ ions are negligible, 
consistent with the proposal by Karmakar ${et}$ ${al}$ \cite{Karmakar}. 
The ferromagnetic components obtained by subtracting the paramagnetic components from 
the XMCD spectrum at $H$ = 0.5 T is shown in Fig. \ref{MDXMCD}(d). 
From the line-shape analysis, we conclude that the ferromagnetic components are originated from both 
predominant Fe$^{3+}$ and a small amount of Fe$^{2+}$ ions, 
where the composition ratios of the Fe$^{3+}$ ($O$$_h$), Fe$^{3+}$ ($T$$_d$) and Fe$^{2+}$ ($T$$_d$) are 
about $\sim$ 60 \%, $\sim$ 25 \% and $\sim$ 15 \%, respectively. 
In Fig. \ref{MDXMCD}(d), peaks due to the Fe$^{3+}$ ions at the $T$$_d$ and $O$$_h$ sites occur in the opposite directions. 
This clearly implies the presence of Fe$^{3+}$ ($T$$_d$)-Fe$^{3+}$ ($O$$_h$) antiferromagnetic coupling. 
Therefore, it is possible that this sample exhibits "weak ferrimagnetism" due to the Fe$^{3+}$ ions 
occupying two sets of nonequivalent positions ($T$$_d$ and $O$$_h$ sites) in unequal numbers and 
in antiparallel configurations so that there is a net moment \cite{Gilleo}. 
That is, the ferromagnetism is caused by the difference in the electron numbers 
between up and down spins at $T$$_d$ ($O$$_h$) and $O$$_h$ ($T$$_d$) sites. 
Indeed, the room temperature ferromagnetism of the ferrite $\gamma$-Fe$_2$O$_3$ due to 
such Fe$^{3+}$-Fe$^{3+}$ antiferromagnetic coupling has been reported \cite{Boa}. 
In addition to this, Fe$^{2+}$-Fe$^{2+}$, Fe$^{3+}$-Fe$^{3+}$ exchange interactions and 
Fe$^{3+}$-Fe$^{2+}$ double exchange interaction are considered to exist. 

\begin{figure}
\includegraphics[width=8.0cm]{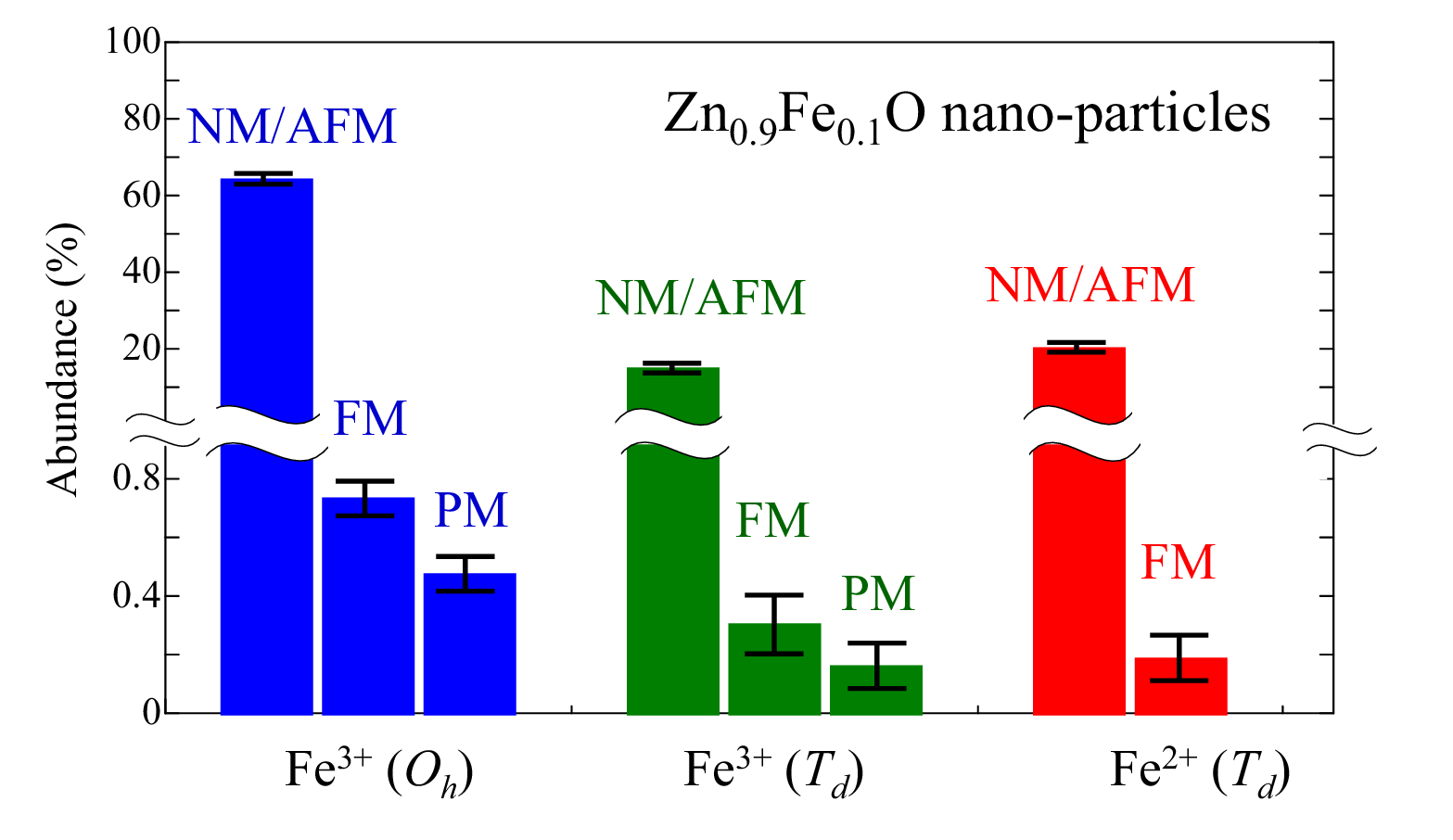}
\caption{\label{Abundance}(Color online) Relative abundance of the nonmagnetic/antiferromagnetic (NM/AFM), 
ferromagnetic (FM) and paramagnetic (PM) components of the Fe 
ions [Fe$^{3+}$ ($O$$_h$), Fe$^{3+}$ ($T$$_d$) and Fe$^{2+}$ ($T$$_d$)] 
in the Zn$_{0.9}$Fe$_{0.1}$O nano-particles.}
\end{figure}

From the experimental and cluster-model calculation results, we have estimated the relative abundance of the nonmagnetic/antiferromagnetic (NM/AFM), 
ferromagnetic (FM) and paramagnetic (PM) components of the Fe ions [Fe$^{3+}$ ($O$$_h$), Fe$^{3+}$ ($T$$_d$) and Fe$^{2+}$ ($T$$_d$)] 
as shown in Fig. \ref{Abundance}. 
The NM/AFM components are originated from the strongly antiferromagnetic coupled Fe ions and are dominant in this sample. 
On the other hand, the weak ferromagnetic and paramagnetic components are originated mainly from Fe$^{3+}$ as well as 
from a small amount of Fe$^{2+}$ ions and uncoupled Fe$^{3+}$ ions, respectively. 
In addition, our analysis confirms the predominance of the Fe$^{3+}$ ($O$$_h$ and $T$$_d$) ions in this sample. 
Considering that the Fe$^{3+}$ ions are dominant in the surface region of the nano-particles probed by surface-sensitive Fe 3$p$-3$d$ RPES, 
the presence of the Fe$^{3+}$ must be a surface effects. 
If Zn vacancies, which dope the system with holes, 
are present near Fe$^{2+}$ ($T$$_d$) ions substituting Zn sites, 
the Fe$^{2+}$ ($T$$_d$) ions will be converted to Fe$^{3+}$ ($T$$_d$) \cite{Karmakar}. 
This will occur mostly in the surface region of the nano-particles, 
where the probability of the presence of vacancies is higher. 
The Fe ions should be at the $T$$_d$ sites if Fe simply substitutes for Zn site in the Fe-doped ZnO system. 
One possible origin for the presence of the Fe ($O$$_h$) ions is due to the interstitial impurities. 
However, Karmakar ${et}$ ${al}$. have performed EPR, x-ray diffraction (XRD) and M\"ossbauer measurements 
on the same samples and the presence of interstitial impurities has been excluded \cite{Karmakar}. 
The other possible origin for the presence of the Fe ($O$$_h$) ions 
is due to excess oxygen at the surface of the nano-particles. 
According to the literature about the molecular dynamics simulations of Fe$_2$O$_3$ nano-particles \cite{B.T.H.L.Khanh}, 
to achieve local charge neutrality, it is expected that oxygen atoms have a tendency to concentrate on the surface of 
the Zn$_{0.9}$Fe$_{0.1}$O nano-particles. 
Due to the excess oxygen, the Fe ions in the surface region of the Zn$_{0.9}$Fe$_{0.1}$O nano-particles 
are coordinated to a larger number of oxygen atoms as if they were at the $O$$_h$ sites. 
Indeed, Chen ${et}$ ${al}$. \cite{LXChen} have reported that 
Fe ($T$$_d$) ions in the surface region of Fe$_3$O$_4$ nano-particles have a tendency to be converted to Fe ($O$$_h$). 
In the nano-particle systems, the surface modification may dramatically affect the electronic structure and magnetic properties.

\section{Conclusion}

In summary, we have performed XPS, RPES, XAS and XMCD measurements on Zn$_{0.9}$Fe$_{0.1}$O nano-particles, 
which exhibit ferromagnetism at room temperature. 
From the experimental and cluster-model calculation results, 
we find that Fe atoms are predominantly in the Fe$^{3+}$ ionic state with mixture of a small amount of Fe$^{2+}$ and 
that Fe$^{3+}$ ions are dominant in the surface region of the nano-particles. 
It is shown that the room temperature ferromagnetism in the Zn$_{0.9}$Fe$_{0.1}$O nano-particles 
is primarily originated from the antiferromagnetic coupling 
between unequal amounts of Fe$^{3+}$ ions occupying two sets of nonequivalent 
positions in the region of the XMCD probing depth of $\sim$ 2-3 nm.

\section*{Acknowledgement}
We thank T. Koide and D. Asakura for useful discussion and comments. 
We thank T. Okuda and A. Harasawa for their valuable technical support for the experiment at PF. 
The experiment at SPring-8 was performed under the approval of the Japan Synchrotron Radiation Research Institute (JASRI) (proposal no. 2007B3825). 
The experiment at PF was approved by the Photon Factory Program Advisory Committee (Proposal No. 2006G002). 
This work was supported by a Grant-in-Aid for Scientific Research in Priority Area 
"Creation and Control of Spin Current" (19048012) from MEXT, 
Japan and a Global COE Program "the Physical Sciences Frontier", from MEXT, Japan and 
an Indo-Japan Joint Research Project "Novel Magnetic Oxide Nano-Materials Investigated by Spectroscopy and 
${ab}$-initio Theories" from JSPS, Japan.

\end{document}